\documentclass[conference,letterpaper]{IEEEtran}
\usepackage[utf8]{inputenc} 
\usepackage[T1]{fontenc}
\usepackage{url}

\usepackage[cmex10]{amsmath}
\usepackage{amsmath}
\usepackage{cuted}
\allowdisplaybreaks
\usepackage{amsmath,amssymb,amsfonts}
\usepackage{algorithmicx}
\usepackage{algorithm,algpseudocode}
\usepackage{algcompatible}
\usepackage{array}
\usepackage[caption=false,font=normalsize,labelfont=sf,textfont=sf]{subfig}
\usepackage{textcomp}

\usepackage{url}
\usepackage{verbatim}
\usepackage{cuted}
\usepackage{dblfloatfix} 
\usepackage{mathrsfs}
\usepackage{caption}
\usepackage{cite}
\usepackage{bm}
\usepackage{xcolor}
\usepackage{colortbl}
\usepackage{multirow}
\usepackage{bbm}
\usepackage[figuresright]{rotating}
\allowdisplaybreaks
\newtheorem{example}{Example}

\newtheorem{definition}{Definition}
\newtheorem{theorem}{Theorem}
\newtheorem{lemma}{Lemma}

\ifodd 1

\else

\fi

\begin{document}
\title{Fundamental Limits of Coded \\ Polynomial Aggregation} 
\author{
\IEEEauthorblockN{Xi Zhong\textsuperscript{1}, Jörg Kliewer\textsuperscript{2} and Mingyue Ji\textsuperscript{1}}

\IEEEauthorblockA{\textit{\textsuperscript{1}Department of Electrical and Computer Engineering}, \textit{University of Florida}, Gainesville, FL, USA\\
Email: \{xi.zhong, mingyueji\}@ufl.edu }

\IEEEauthorblockA{\textit{\textsuperscript{2}Department of Electrical and Computer Engineering}, \textit{New Jersey Institute of Technology}, Newark, NJ, USA\\
Email: jkliewer@njit.edu }}

\maketitle

\begin{abstract}
We introduce the problem of \emph{coded polynomial aggregation} (CPA) in distributed computing systems, where the goal is to recover a weighted sum of polynomial computations using the responses from distributed workers.
A fundamental question in CPA is the minimum number of responses required to exactly recover the computation.
A straightforward strategy is individual decoding, where the system first decodes all individual polynomial computations and then aggregates them, which requires at least $d(K-1)+1$ responses for a degree-$d$ polynomial computed on $K$ datasets.
In this paper, we study CPA in the regime where the number of responses satisfies $N \le d(K-1)$, and show that exact recovery can be achieved by directly exploiting the aggregation structure and the algebraic properties of polynomials.
In particular,  we establish a necessary and sufficient condition for exact recovery in CPA.
Moreover, we characterize the minimum number of responses, showing that it equals $\left\lfloor \frac{K-1}{2} \right\rfloor + 1$ when $d=1$, and $(d-1)(K-1)+1$ when $d \ge 2$.  
These thresholds are strictly smaller than that required by individual decoding.
We further provide explicit constructions of CPA schemes that achieve this bound.
\end{abstract}

\section{Introduction}
Coded distributed computing is a fundamental paradigm for executing large-scale data processing tasks, where a master node partitions a computation task, assigns sub-tasks to multiple workers, and aggregates their responses to recover the desired computation. 
One commonly used metric in coded distributed computing systems is the minimum amount of information that must be collected from the workers to recover the computation. This metric can be characterized by the \emph{minimum number of responses}, i.e., the smallest number of worker responses required for exact recovery.

For matrix multiplications and polynomial computations, many prior works apply polynomial-based coded computing techniques. 
Specifically, the authors in \cite{8002642} use one-dimensional maximum distance separable (MDS) codes to encode matrices, achieving a minimum number of responses, which equals the partition level of the data matrix. Product codes \cite{8006963} use two-dimensional MDS codes and apply iterative decoding algorithms. Polynomial codes \cite{yu2018polynomialcodesoptimaldesign} encode sub-matrices as coefficients of a univariate polynomial and interpolate this polynomial using responses from workers. PolyDot codes and MatDot codes \cite{8765375} extend the interpolation-based approach to trade off computation and communication costs by partitioning datasets. Entangled polynomial codes \cite{9174167, 8949560} exploit a bilinear structure and extend polynomials codes to scenarios such as secure, private, and batch computation. To construct numerically stable interpolation schemes, the authors in \cite{8849468} apply Chebyshev polynomials to coded computing. Instead of univariate polynomials, bivariate polynomial coded schemes \cite{9322629} develop two-dimensional interpolation, enabling finer trade-offs among storage, computation, and communication.
Differing from encoding sub-computations as coefficients of a polynomial, Lagrange coded computing \cite{yu2019lagrangecodedcomputingoptimal} encodes sub-computations as evaluations of Lagrange polynomials, achieving a lower decoding complexity.
Apart from decoding exact computations, several works consider approximated coded computing for arbitrary functions under numerical or probabilistic guarantees \cite{9713954, moradi2024codedcomputingresilientdistributed, 11195640}.

A common paradigm underlying many existing coded computing schemes is \emph{individual decoding}, where the master reconstructs all individual sub-computations before computing the desired output.
In particular, for matrix multiplications, the works in \cite{8002642, 8006963, yu2018polynomialcodesoptimaldesign, 8765375, 9174167, 8949560, 8849468,9322629} decode all sub-matrices contained in the desired matrix, and therefore recover the computation.
However, for aggregation tasks, individual decoding is sufficient for exact recovery but may lead to redundant computations.
To compute aggregation tasks in the form of gradient computations, \cite{9488815} proposed gradient coding, which introduces redundancy into gradient computations to mitigate slow  workers.
A different class of aggregation schemes considers linearly separable computations \cite{9614153}, where the authors focus on computing weighted aggregations of arbitrary computation functions with minimizing the communication cost.

Despite these extensive efforts, the problem of computing a weighted aggregation of polynomial computations has not been studied as a standalone objective.
Specifically, we consider the computation of a weighted aggregation, 
\begin{equation}
\label{eq-goal}
\boldsymbol{Y} \triangleq  \sum_{k=0}^{K-1} w_{k}\, F(\boldsymbol{X}_k),
\end{equation}
where $w_k \in \mathbb{C}$ are non-zero weights, and $F(\cdot)$ is a polynomial of degree $d$ that operates element-wise on each data matrix $\boldsymbol{X}_k \in \mathbb{C}^{q \times v}$.
This formulation commonly arises in distributed learning and signal processing, where the objective is to recover an aggregated output rather than all individual computations.
A straightforward strategy is  individual decoding, where workers compute polynomial computations on encoded data, and the master interpolates the resulting polynomial to recover all $F(\boldsymbol{X}_k)$ for $k \in [K]$.
Under individual decoding, the minimum number of responses for exact recovery is $d(K-1)+1$.
However, the objective is to recover a weighted aggregation rather than each $F(\boldsymbol{X}_k)$, which leads to the following questions.
\begin{enumerate}
\item \emph{Can the structures of aggregation and polynomials be exploited to reduce the number of worker responses for exact recovery?}
\item \emph{What is the minimum number of responses?}
\item \emph{Do there exist explicit schemes that achieve this minimum number of responses?}
\end{enumerate}
In this paper, we address these questions by characterizing the fundamental limits of coded polynomial aggregation (CPA) and constructing optimal schemes, as summarized below.
\begin{enumerate}
\item We propose a CPA framework for computing the weighted aggregation in \eqref{eq-goal},
and establish a necessary and sufficient condition for exact recovery
in the regime $N \le d(K-1)$, where individual decoding is infeasible.
\item We characterize the minimum number of responses $N^*$ required for a feasible CPA scheme.
\item We provide explicit schemes that achieve this bound $N^*$.
\end{enumerate}

\section{Problem Formulation}
We consider a distributed computing system consisting of a master node and a set of $N$ worker nodes, indexed by $[N] \triangleq \{0,1,\ldots, N-1\}$.
Given $K$ data matrices $\boldsymbol{X}_k \in \mathbb{C}^{q \times v}$ for $k \in [K]$, a polynomial function $F(\cdot)$ of degree $d$  that operates element-wise on each data matrix, and a weight vector $\boldsymbol{w} \in \mathbb{C}^{K}$, where each element $w_{k} \neq 0$ for $k \in [K]$, the objective of the system is to compute the weighted aggregation defined in \eqref{eq-goal}, using the responses from workers. Each worker returns one response to the master.

\subsection{Proposed CPA}
A CPA scheme consists of three phases: encoding, computing, and decoding.

\subsubsection{Encoding}
The master selects a set of $K$ distinct \emph{data points} $\{\alpha_k \in \mathbb{C} : k \in [K]\}$, and interpolates an encoder polynomial $E(z)$ such that $E(\alpha_k) =$ $ \boldsymbol{X}_k$ for all $k \in [K]$.
Next, the master selects a set of $N$ distinct \emph{evaluation points} 
$\{\beta_n$ $\in \mathbb{C} :$ $n \in [N]\}$ satisfying
$\{\alpha_k :$ $k \in [K]\}$ $\cap$ $\{\beta_n :$ $n \in [N]\}$ $=$ $\emptyset$.
The master evaluates $E(z)$ at $\{\beta_n: n \in [N]\}$ and sends the coded matrix
$E(\beta_n)$ to worker $n$.

\subsubsection{Computing}
Each worker $n \in [N]$ computes $F(E(\beta_n))$ locally and returns the result to
the master.

\subsubsection{Decoding}
Upon receiving responses from all  $N$ workers, the master interpolates a decoder polynomial $D(z)$  such that
$D(\beta_n) = F(E(\beta_n))$ for all $n \in [N]$. The master then evaluates $D(z)$ at the data points
$\{\alpha_k : k \in [K]\}$ and obtains
\begin{equation}
\label{eq-decoded-output}
\widehat{\boldsymbol{Y}}
\triangleq
\sum_{k=0}^{K-1} w_k\, D(\alpha_k).
\end{equation}

We define the feasibility of a CPA scheme as follows.
\begin{definition}[Feasibility of CPA]
Fix positive integers $K$, $d$, and $N$, data points $\{\alpha_k$ $:$ $k \in [K]\}$ and evaluation points $\{\beta_n : n \in [N]\}$. A CPA scheme is  \emph{feasible} if  $\widehat{\boldsymbol{Y}} = \boldsymbol{Y}$.
\hfill $\diamond$
\end{definition}

In this paper, we treat the data points $\{\alpha_k : k \in [K]\}$ as fixed system parameters\footnotemark.
\footnotetext{Allowing joint design of the data points $\{\alpha_k : k \in [K]\}$ and the evaluation points $\{\beta_n : n \in [N]\}$ may further enlarge the feasible design space of CPA schemes.}
We assume that they are pairwise distinct, i.e., $\alpha_i \neq \alpha_j$ for all $i \neq j$, and generic\footnotemark in the sense that they are chosen outside a proper algebraic variety determined by the system parameters $K$, $N$, $d$ and $\boldsymbol{w}$.
\footnotetext{For background on relevant concepts in algebraic geometry, such as genericity, algebraic varieties, Lebesgue
measure zero, and Zariski-open dense subsets used in this paper, we refer the reader to \cite{Shafarevich1995}.}

We define the minimum number of responses as follows.
\begin{definition}[Minimum Number of Responses]
For positive integers $K$ and $d$, and a given generic set of pairwise distinct data points
$\{\alpha_k : k \in [K]\}$, the \emph{minimum number of responses} is defined as the smallest
integer $N$ such that there exists a feasible CPA scheme.
\hfill $\diamond$
\end{definition}

In this paper, the goal is to characterize the minimum number of responses and to construct explicit schemes that achieve this bound.

\subsection{Individual Decoding Baseline}
Existing results on polynomial codes, including
\cite{yu2018polynomialcodesoptimaldesign, 8765375, 9174167, 8949560, 8849468, 9322629, yu2019lagrangecodedcomputingoptimal},
recover the desired computation by decoding all individual sub-computations through reconstructing polynomial evaluations or coefficients.
Among these works, we set \emph{Lagrange coded computing}~\cite{yu2019lagrangecodedcomputingoptimal} as the baseline to our setting.
It focuses on polynomial computations and encodes each data matrix $\boldsymbol{X}_k$ as an evaluation of an encoder polynomial $E(z)$.
When applied to the CPA setting, Lagrange coded computing leads to the following CPA based on individual decoding.
\begin{definition} [CPA Based on Individual Decoding]
    We define \emph{CPA based on individual decoding} as a decoding strategy, where
    the master reconstructs all individual evaluations $F(\boldsymbol{X}_k)$ by
    interpolating a polynomial $D(z)$ satisfying $F(\boldsymbol{X}_k) = D(\alpha_k)$
    for all $k \in [K]$. 
    \hfill$\diamond$
\end{definition}

\begin{figure*}
\centering
\includegraphics[width=0.85\textwidth]{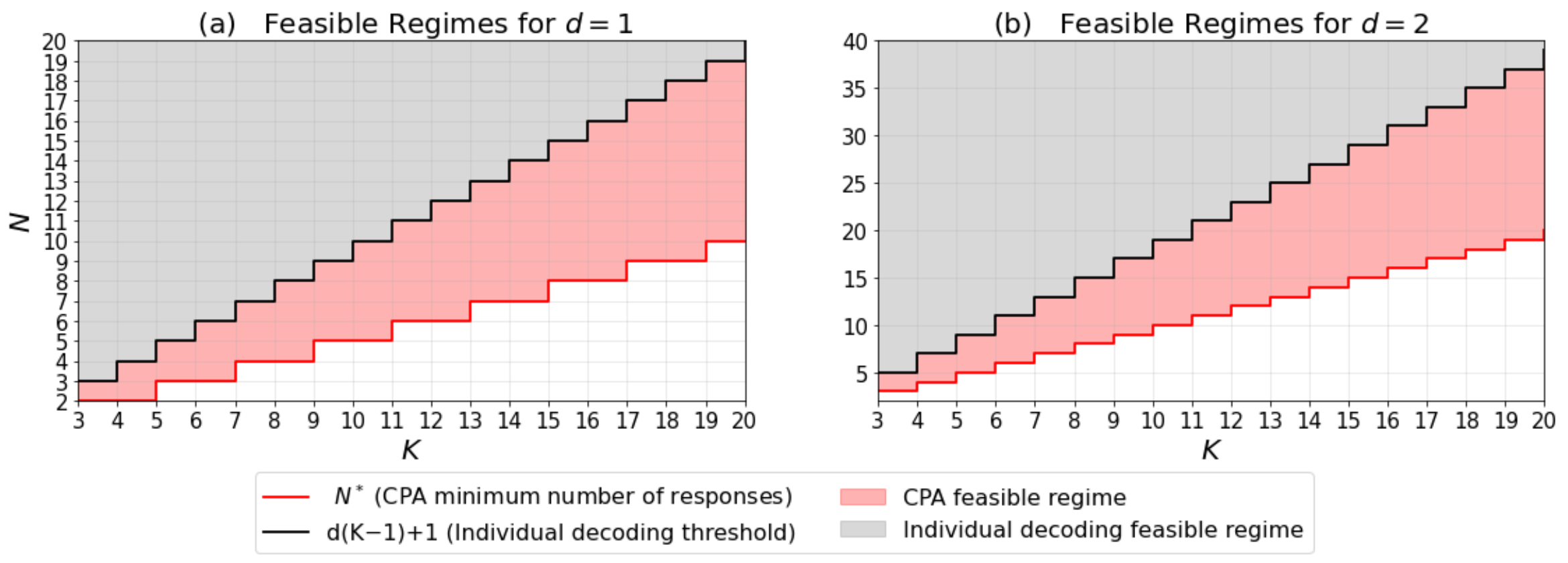}
\caption{Feasible regimes of CPA and the baseline CPA scheme based on individual decoding for computation degrees $d=1$ and $d=2$.
For a fixed $K$, the feasible regime consists of the values of $N$ for which the corresponding scheme is feasible.
The red region corresponds to CPA, which enables feasibility with fewer responses ($N \le d(K-1)$) via the orthogonality conditions in Theorem~\ref{th-condition}, with the minimum number of responses $N^*$ characterized in Theorem~\ref{th-bound-case1}.
The gray region corresponds to the baseline scheme based on individual decoding, which is feasible only when $N \ge d(K-1)+1$, as stated in Lemma~\ref{le-local}.}
\label{fig:feasible_regimes}
\end{figure*}

From standard polynomial interpolation arguments, the following lemma characterizes the minimum number of responses for  CPA  based on individual decoding.
\begin{lemma}
\label{le-local}
For integers $K$ and $d$, a CPA scheme based on individual decoding is feasible if and only if $N \ge d(K-1)+1$.
\end{lemma}


In this paper, we study CPA in the regime $N \le d(K-1)$, where a CPA scheme based on individual decoding is infeasible.
We show that exact recovery is achievable  by directly exploiting the aggregation structure and the algebraic
properties of polynomials.

\section{Main Results}
\subsection{Necessary and Sufficient Orthogonality Conditions}
We characterize a necessary and sufficient condition for the existence of a feasible CPA scheme when $N \le d(K-1)$.

\begin{theorem}
\label{th-condition}
For positive integers $K$, $d$, and $N$ with $2 \le N \le d(K-1)$,
let $C \triangleq d(K-1)-N+1$.
A CPA scheme is feasible if and only if the data points $\{\alpha_k \in \mathbb{C} : k \in [K]\}$
and evaluation points $\{\beta_n \in \mathbb{C} : n \in [N]\}$
satisfy $\{\alpha_k : k \in [K]\} \cap \{\beta_n : n \in [N]\} = \emptyset$,
and
\begin{equation}
\label{eq-equations}
\sum_{k=0}^{K-1} w_k \, P(\alpha_k)\, \alpha_k^j = 0,
\quad \forall j \in [C],
\end{equation}
where
\begin{equation}
\label{eq-partial-P}
P(z) \triangleq \prod_{n \in [N]} (z - \beta_n).
\end{equation}
\end{theorem}
\begin{IEEEproof}
The proof is provided in Section~\ref{sec-appen-1}.
\end{IEEEproof}

Theorem~\ref{th-condition} shows that a CPA scheme is feasible if and only if the data points $\{\alpha_k: k\in [K]\}$ and evaluation points $\{\beta_n: n\in [N]\}$ satisfy a system of orthogonality conditions, as shown in \eqref{eq-equations}.
The intuition behind the proposed conditions in \eqref{eq-equations} is as follows.
When $N \le d(K-1)$, the decoder polynomial $D(z)$ has a strictly smaller degree than the polynomial $F(E(z))$, and therefore cannot recover the individual evaluations $F(E(\alpha_k))$.
However, the recovery error $\widehat{\boldsymbol{Y}}-\boldsymbol{Y}$ can be expressed as a linear combination of   $\sum_{k=0}^{K-1} w_k P(\alpha_k)\alpha_k^j$ for $j \in [C]$. The orthogonality conditions in \eqref{eq-equations} force all such quantities to be zero, thereby eliminating the recovery error and enabling exact recovery of $\boldsymbol{Y}$. 


\subsection{Minimum Number of Responses}

The minimum number of responses of CPA is characterized as follows.
\begin{theorem}
\label{th-bound-case1}
For positive integers $K$, $d$, and $N$ with $2 \le N \le d(K-1)$, and for a given generic set of pairwise distinct data points $\{\alpha_k : k \in [K] \}$,
there exist evaluation points $\{\beta_n : n \in [N]\}$ such that CPA is feasible
if and only if $N \ge N^*$, where
\begin{equation}
\label{eq-dk_bound}
N^* =
\begin{cases}
\left\lfloor \dfrac{K-1}{2} \right\rfloor +1, & \text{if } d = 1, \\[6pt]
(d-1)(K-1) +1, & \text{if } d \geq 2.
\end{cases}
\end{equation}
\end{theorem}

From Theorem~\ref{th-bound-case1}, when $d=1$, the minimum number of responses of CPA scales to approximately  half of that required by individual decoding as $K$ grows large.
When $d \ge 2$, CPA reduces the required number of responses by $K-1$ compared to the individual decoding baseline in Lemma~\ref{le-local}.

Fig.~\ref{fig:feasible_regimes} illustrates the feasible regimes and the minimum number of responses, required by CPA and the baseline CPA scheme based on individual decoding.  
From Fig.~\ref{fig:feasible_regimes}, for both $d=1$ and $d=2$, CPA has a strictly smaller minimum number of responses for any $K$. Moreover, the difference between the proposed CPA schemes and the baseline CPA scheme based on individual decoding grows with $K$, indicating that the benefit of CPA becomes more pronounced as the number of data matrices increases.

\begin{algorithm}[H]
\caption{Construction of $\{\beta_n : n \in [N]\}$}
\label{al-beta}
\begin{algorithmic}[1]
\REQUIRE $K, d, N$, $\boldsymbol{w}$, $\{\alpha_k : k \in [K]\}$
\STATE $C \gets d(K-1) - N + 1$
\STATE Construct $\boldsymbol{V} \in \mathbb{C}^{C \times K}$ with
$V[j,k] = \alpha_k^{\,j}$ for $j \in [C]$, $k \in [K]$
\STATE Construct $\boldsymbol{A} \in \mathbb{C}^{K \times (N+1)}$ with
$A[k,n] = \alpha_k^{\,n}$ for $k \in [K]$, $n \in [N+1]$
\STATE $\boldsymbol{U} \gets \boldsymbol{V}\,\operatorname{diag}(\boldsymbol{w})\,\boldsymbol{A}$
\STATE Select a nonzero vector
$\boldsymbol{c} = (c_0,\ldots,c_N)^{\mathsf T} \in \ker(\boldsymbol{U})$
\STATE Construct the polynomial
$P_{\text{form}}(z) = \sum_{n=0}^{N} c_n z^n$
\STATE Compute the $N$ roots $\{\beta_n : n \in [N]\}$ of $P_{\text{form}}(z)$
\STATE \textbf{return} $\{\beta_n : n \in [N]\}$
\end{algorithmic}
\end{algorithm}

\subsection{Achievable CPA Schemes}
We next present an achievable CPA scheme by constructing evaluation points $\{\beta_n : n \in [N]\}$ for a given generic set of pairwise distinct data points $\{\alpha_k : k \in [K]\}$.
\begin{theorem}
\label{th-algorithm}
For positive integers $K$, $d$, and $N$ satisfying $N^* \le N \le d(K-1)$, and for a given generic set of pairwise distinct data points $\{\alpha_k : k \in [K]\}$,
Algorithm~\ref{al-beta} constructs evaluation points $\{\beta_n : n \in [N]\}$ that satisfy the conditions in Theorem~\ref{th-condition}.
\end{theorem}
\begin{IEEEproof}
The correctness of Algorithm~\ref{al-beta} follows directly from the constructive sufficiency proof of Theorem~\ref{th-bound-case1}.
Importantly, all but a Lebesgue measure-zero subset of vectors $\boldsymbol{c} \in \ker(\boldsymbol{U})$ lead to polynomials $P_{\text{form}}(z)$ such that its roots $\{\beta_n : n \in [N]\}$ are distinct and satisfy the orthogonality conditions in \eqref{eq-equations} and $\{\beta_n: n\in [N]\} \cap \{\alpha_k: k\in [K]\} = \emptyset$.
Details are provided in Section~\ref{sec-2-suff}.
\end{IEEEproof}

The following example illustrates the construction in Algorithm \ref{al-beta} for the case $d=1$.
\begin{example}
Consider the linear case $d=1$ with $K=3$ and
data points $\{\alpha_0$, $\alpha_1$, $\alpha_2\}$ $=$ $\{-1$, $0$, $1\}$,
and weights $\boldsymbol{w}=(-0.5$, $1$, $0.5)$.
Under individual decoding, $d(K-1)$ $+1$ $=3$ workers are required, and the evaluation points
$\{\beta_0$, $\beta_1$, $\beta_2\}$ can be chosen arbitrarily as long as they are distinct and disjoint from $\{-1$, $0$, $1\}$.

In contrast, CPA allows $N=2$, which gives $C=1$.
Applying Algorithm~\ref{al-beta}, we obtain
$\boldsymbol{V}=\begin{bmatrix}1 & 1 & 1\end{bmatrix} \text{ and }
\boldsymbol{A}=
\begin{bmatrix}
1 & -1 & 1 \\
1 & 0  & 0 \\
1 & 1  & 1
\end{bmatrix}$,
which yields
$\boldsymbol{U}=\begin{bmatrix}1 & 1 & 0\end{bmatrix}$.
Solving $\boldsymbol{U}\boldsymbol{c}=\boldsymbol{0}$ gives
$\boldsymbol{c}$ $=(-0.0506$, $0.0506$, $0.5)$ and $P_{\text{form}}(z)$ $=-0.0506$ $+ 0.0506z$ $+ 0.5z^2$.
The two roots are $\beta_0 = - 0.37272$ and $\beta_1 = 0.27152$, which are distinct and disjoint from $\{-1,0,1\}$.
Finally, the orthogonality conditions in Theorem~\ref{th-condition} are numerically satisfied, with
$\sum_{k=0}^{2} w_k P(\alpha_k) \alpha_{k}^{0}\approx 5.5\times10^{-17}$.
\end{example}

\section{Proof of Theorem~\ref{th-condition}}
\label{sec-appen-1}
We first consider the scalar case where $\boldsymbol{X}_k$ reduces to a scalar $x_k$ and the desired output $\boldsymbol{Y}$ reduces to  $y$. The extension to the matrix case follows directly by applying the same argument element-wise, since the polynomial $F(\cdot)$ operates element-wise on the data matrices.
Define the error polynomial $\Delta(z) \triangleq D(z) - F(E(z))$, which satisfies $\deg(\Delta)$ $ \le$ $\max\{\deg(D)$, $\deg(F(E))\}$ $\le$ $\max\{N-1$, $d(K-1)\} = d(K-1)$.
The recovery error $\widehat{y}-y$ can be expressed by
\begin{equation}
\label{eq-pf-gmse}
\widehat{y}-y = \sum_{k=0}^{K-1} w_k \Delta(\alpha_k),
\end{equation}
as $\widehat{y}-y$ $=$ $\sum_{k=0}^{K-1} w_k\big(D(\alpha_k)-F(x_k)\big)$ $\overset{(a)}{=}$ $\sum_{k=0}^{K-1} w_k\big(D(\alpha_k)-F(E(\alpha_k))\big)$ $=$ $\sum_{k=0}^{K-1} w_k \Delta(\alpha_k)$, where $(a)$ is due to $x_k = E(\alpha_k)$.
During decoding, $D(\beta_n)=F(E(\beta_n))$  imply $\Delta(\beta_n)$ $=$ $0$ for all $n\in[N]$.
Consequently, we can write $\Delta(z)$ as 
\begin{equation}
\label{eq-re-delta-poly}
\Delta(z) = \prod_{n\in[N]}(z-\beta_n)\cdot R(z) = P(z)\cdot R(z),
\end{equation}
where $P(z)=\prod_{n\in[N]}(z-\beta_n)$ has degree $N$, as shown in \eqref{eq-partial-P}, and $R(z)$ is a polynomial satisfying $\deg(R)$ $=$ $\deg(\Delta)-\deg(P)$ $\le$ $d(K-1)-N$.
Writing $R(z)$ $=$ $\sum_{j=0}^{\deg(R)} r_j z^j$, where $\{r_j:$ $j \in [\deg(R)+1]\}$ are arbitrary complex coefficients.
Substituting \eqref{eq-re-delta-poly} into \eqref{eq-pf-gmse} yields
\begin{align*}
    \widehat{y}-y = \sum_{j=0}^{\deg(R)} r_j  \sum_{k=0}^{K-1} w_k P(\alpha_k)\alpha_k^j ,
\end{align*}
as $ \widehat{y}-y$ $=$ $\sum_{k=0}^{K-1}$ $w_k P(\alpha_k) R(\alpha_k)$ $=$ $\sum_{k=0}^{K-1}$ $w_k P(\alpha_k)$ $\sum_{j=0}^{\deg(R)} r_j \alpha_k^j$ $= \sum_{j=0}^{\deg(R)} r_j ( \sum_{k=0}^{K-1} w_k P(\alpha_k)\alpha_k^j )$.
Therefore, $\widehat{y}-y=0$ for all admissible choices of $R(z)$ if and only if $\sum_{k=0}^{K-1} w_k P(\alpha_k)\alpha_k^j = 0$ for all $j=$ $0$, $1$, $\ldots$, $d(K-1)-N$. This establishes the necessary and sufficient condition for feasibility in the scalar case.
The same argument applies element-wise to the matrix-valued case, and hence the condition remains necessary and sufficient for feasibility.
This completes the proof of Theorem~\ref{th-condition}.

\section{Proof of Theorem~\ref{th-bound-case1} }
\label{sec-sppen-2}
To prove Theorem~\ref{th-bound-case1}, it is equivalent to show that there exists a set $\{\beta_n : n \in [N]\}$ satisfying the following three properties:
\begin{align}
& \textbf{(Orthogonality): }  
\sum_{k=0}^{K-1} w_{k}\,P(\alpha_k)\,\alpha_k^j = 0,\quad j\in [C]. \label{eq-proof-equations} \\
& \textbf{(Disjointness): } \{\alpha_k : k\in[K]\}\cap\{\beta_n : n\in[N]\}=\emptyset. \label{eq-disjoint} \\
& \textbf{(Distinction): }
\beta_n\neq \beta_{n'} \text{ for all } n\neq n'. \label{eq-distinction}
\end{align}
if and only if $N\ge N^*$, where $N^*$ is given in \eqref{eq-dk_bound}. 

\subsection{Reformulation of Orthogonality}
Let $\operatorname{diag}(\boldsymbol{w})$ denote the diagonal matrix with diagonal entries given by $\boldsymbol{w}$.
Define $\boldsymbol{p}=(p_0,\ldots,p_{K-1})^{\mathsf T}\in\mathbb{C}^{K}$ with
$p_k\triangleq P(\alpha_k)$.
Let $\boldsymbol{V}\in\mathbb{C}^{C\times K}$ be the Vandermonde matrix with entries $V[j,k]=\alpha_k^j$ for $j\in[C]$ and $k\in[K]$.
Substituting $\boldsymbol{V}$ and $\boldsymbol{p}$ into \eqref{eq-proof-equations} yields
\begin{equation}
\label{eq-linear_system}
\sum_{k=0}^{K-1} w_{k}\,P(\alpha_k)\,\alpha_k^j = 0, j\in[C]
\ \Leftrightarrow\ 
\boldsymbol{V}\,\operatorname{diag}(\boldsymbol{w})\,\boldsymbol{p}=\boldsymbol{0}.
\end{equation}

Next, we express $\boldsymbol{p}$ in terms of the coefficients of the polynomial $P(z)$.
Let $P_{\mathrm{form}}(z)=c_{0}+c_{1}z+\ldots+c_{N}z^{N}$ be a polynomial with distinct roots $\{\beta_n:n\in[N]\}$, where $c_{N}\neq 0$.
Equivalently, $P_{\mathrm{form}}(z)=c_{N}(z-\beta_{0})(z-\beta_{1})\ldots(z-\beta_{N-1})$.
From \eqref{eq-partial-P}, $P(z)=\frac{1}{c_{N}}P_{\mathrm{form}}(z)$ for some coefficient vector $\boldsymbol{c}=(c_0,\ldots,c_N)^{\mathsf T}\in\mathbb{C}^{N+1}$ with $c_{N}\neq 0$.
Then,
\begin{equation}
\label{eq-pIP}
\boldsymbol{p}=
\begin{bmatrix}
P(\alpha_{0}) &
P(\alpha_{1}) &
\ldots &
P(\alpha_{K-1})
\end{bmatrix}^{\mathsf T}
=
\frac{1}{c_{N}}\,\boldsymbol{A}\boldsymbol{c},
\end{equation}
where we define $\boldsymbol{A}\in\mathbb{C}^{K\times (N+1)}$ with entries $A[k,n]=\alpha_k^{n}$ for $k\in[K]$ and $n\in[N+1]$.
Substituting \eqref{eq-pIP} into \eqref{eq-linear_system} yields
\begin{equation}
\label{eq-equal-Uc}
\begin{aligned}
\boldsymbol{V}\operatorname{diag}(\boldsymbol{w})\boldsymbol{p}=\boldsymbol{0}
&\overset{(a)}{\Leftrightarrow} 
\boldsymbol{V}\operatorname{diag}(\boldsymbol{w})\boldsymbol{A}\boldsymbol{c}\cdot\frac{1}{c_{N}}=\boldsymbol{0} \\
&\overset{(b)}{\Leftrightarrow}
\boldsymbol{V}\operatorname{diag}(\boldsymbol{w})\boldsymbol{A}\boldsymbol{c}=\boldsymbol{0} \\
&\overset{(c)}{\Leftrightarrow}
\boldsymbol{U}\boldsymbol{c}=\boldsymbol{0},
\end{aligned}
\end{equation}
where $(a)$ follows from \eqref{eq-pIP}, $(b)$ follows from $c_{N}\neq 0$, and in $(c)$ we define
$\boldsymbol{U}\triangleq\boldsymbol{V}\operatorname{diag}(\boldsymbol{w})\boldsymbol{A}\in\mathbb{C}^{C\times(N+1)}$.

In 
$\boldsymbol{U}\boldsymbol{c}=\boldsymbol{0}$, the vector $\boldsymbol{c}$ corresponds to the coefficients of the polynomial whose roots are $\{\beta_n:n\in[N]\}$.
The matrix $\boldsymbol{U}$ depends only on $\boldsymbol{w}$ and $\{\alpha_k:k\in[K]\}$ and satisfies 
\begin{equation}
\label{eq-rank-U}
\operatorname{rank}(\boldsymbol{U}) \leq  \min(C,K,N+1),
\end{equation}
as $\operatorname{rank}(\boldsymbol{U})$ $=$ $\operatorname{rank} (\boldsymbol{V} \operatorname{diag}(\boldsymbol{w})\boldsymbol{A})$ $\le$ $\min(\operatorname{rank}(\boldsymbol{V}$ $\operatorname{diag}(\boldsymbol{w}))$, $\operatorname{rank}(\boldsymbol{A}))$ $\overset{(a)}{=}$ $\min(\operatorname{rank}(\boldsymbol{V})$, $\operatorname{rank}(\boldsymbol{A}))$ $\overset{(b)}{=}$ $\min(C$, $K$, $N+1)$,
where $(a)$ follows from the fact that $\operatorname{diag}(\boldsymbol{w})$ is invertible. $(b)$ is due to the fact that all $\alpha_k$ are distinct for $k\in [K]$. 

\subsection{Necessity of Theorem~\ref{th-bound-case1}}
Assume that there exist distinct  evaluation points $\{\beta_n: n\in [N]\}$ satisfying the three properties in \eqref{eq-proof-equations}--\eqref{eq-distinction}.

\subsubsection{$C<K$ is necessary}
Assume $C \ge K$.
Then $\operatorname{rank}(\boldsymbol{V}\operatorname{diag}(\boldsymbol{w}))$ $=$ $\operatorname{rank}(\boldsymbol{V})$ $=$ $\min(C$, $K)$ $= K$.
From \eqref{eq-linear_system}, 
$\operatorname{dim}(\operatorname{ker}(\boldsymbol{V}\operatorname{diag}(\boldsymbol{w}))) = K - \operatorname{rank}(\boldsymbol{V}\operatorname{diag}(\boldsymbol{w})) = 0$,
which implies $\boldsymbol{p}=\boldsymbol{0}$, i.e., $P(\alpha_k)=0$ for all $k\in[K]$.
This forces $\{\alpha_k:k \in [K]\}\subseteq \{\beta_n: n\in [N]\}$, contradicting the disjointness condition in \eqref{eq-disjoint}.
Therefore,
\begin{equation}
\label{eq-range-1}
C < K  
\Leftrightarrow N \ge (d-1)(K-1)+1.
\end{equation}

\subsubsection{$C<N+1$ is necessary}
Since there exist non-trivial solutions to $\boldsymbol{U}\boldsymbol{c}=\boldsymbol{0}$ with $\boldsymbol{c}\neq\boldsymbol{0}$,
we must have $\operatorname{dim}(\operatorname{ker}(\boldsymbol{U}))>0$.
Moreover,
$\operatorname{dim}(\operatorname{ker}(\boldsymbol{U})) = (N+1) - \operatorname{rank}(\boldsymbol{U})
\ge (N+1) - \min(C,K,N+1) = (N+1) - \min(C,N+1)$.
Hence, $(N+1) - \min(C,N+1) > 0$, which implies
\begin{equation}
\label{eq-equal-to}
N+1 > C  
\Leftrightarrow N \ge \left\lfloor \frac{d(K-1)}{2} \right\rfloor +1.
\end{equation}
Combining \eqref{eq-range-1} and \eqref{eq-equal-to}, we obtain
$N$ $\geq$ $\max((d-1)(K-1)+1$, $\left\lfloor \frac{d}{2}(K-1) \right \rfloor$ $+1 )$,
which yields \eqref{eq-dk_bound}.
This completes the proof of necessity.

\subsection{Sufficiency of Theorem~\ref{th-bound-case1}}
\label{sec-2-suff}
Assume $N \ge N^*$.
We construct $\{\beta_{n}: n \in [N]\}$ satisfying the conditions in \eqref{eq-proof-equations}-\eqref{eq-distinction}.
The key steps are as follows. 1) We solve the linear system $\boldsymbol{U}\boldsymbol{c} = \boldsymbol{0}$.
2) We select a vector $\boldsymbol{c} \in \ker(\boldsymbol{U})$ with $c_{N}\neq 0$ and obtain the polynomial
$P(z) = \frac{c_0}{c_N} + \frac{c_1}{c_N} z + \ldots + z^N$. From \eqref{eq-equal-Uc}, the orthogonality conditions in \eqref{eq-proof-equations} is satisfied.
3) We compute the $N$ roots of the polynomial $P(z)$ as $\{\beta_n: n \in [N]\}$.
Importantly, we will show that for a generic set of pairwise distinct
$\{\alpha_k:k\in[K]\}$, all but a Lebesgue measure-zero subset of
$\boldsymbol{c}\in\ker(\boldsymbol{U})$ lead to evaluation points
$\{\beta_n:n\in[N]\}$ satisfying
\eqref{eq-disjoint} and \eqref{eq-distinction}.

\subsubsection{Proof of Orthogonality}
Under $N \ge N^*$, we have $\operatorname{dim}(\operatorname{ker}(\boldsymbol{U})) > 0$.
Therefore, there exists a nonzero vector $\boldsymbol{c}\in\ker(\boldsymbol{U})$ such that $\boldsymbol{U}\boldsymbol{c}=\boldsymbol{0}$, as shown in equation $(c)$ of \eqref{eq-equal-Uc}.
Equation $(b)$ in \eqref{eq-equal-Uc} then implies that the polynomial
$P_{\mathrm{form}}(z)=c_{0}+c_{1}z +\ldots+c_{N}z^{N}$
satisfies the orthogonality conditions
$\sum_{k=0}^{K-1} w_{k}\,P_{\mathrm{form}}(\alpha_k)\,\alpha_k^{j}=0$
for all $j\in[C]$.
To ensure that equation $(a)$ in \eqref{eq-equal-Uc} holds, i.e., that the normalized polynomial
$P(z)=\frac{P_{\mathrm{form}}(z)}{c_{N}}$
satisfies the orthogonality conditions in \eqref{eq-proof-equations}, it is required that $c_{N}\neq 0$.
This requirement is captured by the following lemma.
\begin{lemma}
\label{le-C_N}
Define a Zariski open subset of $\ker(\boldsymbol{U})$ as
$\mathcal{C}_{N}\triangleq\{\boldsymbol{c}\in\ker(\boldsymbol{U}) : c_{N}\neq 0\}$.
For generic $\{\alpha_{k}:k\in[K]\}$, $\mathcal{C}_{N} \neq \emptyset$.
\end{lemma}

\begin{IEEEproof}
The proof is sketched as follows.
The set of vectors $\boldsymbol{\alpha}=(\alpha_{0}$, $\ldots$ , $\alpha_{K-1})$ $\in$ $\mathbb{C}^{K}$ for which $\mathcal{C}_{N}$ $=$ $\emptyset$
defines a proper affine variety in $\mathbb{C}^{K}$, and therefore has Lebesgue measure zero.
Hence, for generic $\{\alpha_{k}:k\in[K]\}$,
we have $\mathcal{C}_{N}\neq\emptyset$.
See Appendix A in arXiv \cite{XJM2026CPA} for details.
\end{IEEEproof}

From Lemma~\ref{le-C_N}, for generic $\{\alpha_{k}:k\in[K]\}$,
there exists a vector $\boldsymbol{c}\in\ker(\boldsymbol{U})$
satisfying $c_{N}\neq 0$.
For such a choice of $\boldsymbol{c}$, the polynomial
$P(z)=\frac{c_{0}}{c_{N}}+\frac{c_{1}}{c_{N}}z+\frac{c_{2}}{c_{N}}z^{2}+\ldots+z^{N}$
satisfies \eqref{eq-proof-equations},
and its $N$ roots are the desired   $\{\beta_{n}:n\in[N]\}$.

\subsubsection{Proof of Disjointness}
Condition \eqref{eq-disjoint} is equivalent to requiring
$P(\alpha_k)\neq 0$ for all $k\in[K]$.
Since $P(z)$ and $P_{\mathrm{form}}(z)$ have the same roots,
it suffices to require $P_{\mathrm{form}}(\alpha_k)\neq 0$ for all $k\in[K]$.
From \eqref{eq-pIP}, i.e., $P_{\mathrm{form}}(\alpha_k)=[\boldsymbol{A}\boldsymbol{c}]_k$,
it suffices to show the existence of a vector
$\boldsymbol{c}\in\ker(\boldsymbol{U})$
such that $[\boldsymbol{A}\boldsymbol{c}]_k\neq 0$ for all $k\in[K]$.
This requirement is captured by the following lemma.

\begin{lemma}
\label{le-A_K}
Define a Zariski open subset of $\ker(\boldsymbol{U})$ as $\mathcal{A}_k$ $\triangleq$ $\{\boldsymbol{c}\in\ker(\boldsymbol{U}):$ $[\boldsymbol{A}\boldsymbol{c}]_k$ $\neq$ $0\}$,
and let $\mathcal{C}_{A}$ $\triangleq$ $\bigcap_{k=0}^{K-1}$ $\mathcal{A}_k$.
For generic $\{\alpha_k : k \in [K]\}$, the set $\mathcal{C}_{A}$ is nonempty.
\end{lemma}
\begin{IEEEproof}
The proof is sketched as follows.
 The set of vectors $\boldsymbol{\alpha}$ $=$ $(\alpha_{0}$, $\ldots$, $\alpha_{K-1})$ $\in$ $\mathbb{C}^{K}$
for which $\mathcal{A}_{k}=\emptyset$ defines a proper affine variety in $\mathbb{C}^{K}$, and hence has Lebesgue measure zero.
Hence, for generic $\{\alpha_k: k\in [K]\}$, $\mathcal{A}_k$ is a nonempty Zariski open subset of $\ker(\boldsymbol{U})$.
Since $\ker(\boldsymbol{U})$ is a linear subspace and hence an irreducible affine variety, any finite intersection of nonempty Zariski open subsets of $\ker(\boldsymbol{U})$ is nonempty, i.e., $\mathcal{C}_{A}\neq\emptyset$.
See Appendix B in arXiv \cite{XJM2026CPA} for details.
\end{IEEEproof}

From the above discussion, for generic $\{\alpha_k : k \in [K]\}$, there exists a vector $\boldsymbol{c}\in\ker(\boldsymbol{U})$ such that the roots $\{\beta_n:n\in[N]\}$ of the polynomial $P(z)$ are disjoint from the given set $\{\alpha_k:k\in[K]\}$.
This completes the proof of disjointness.

\subsubsection{Proof of Distinction}
Recall that a polynomial $P(z)$ has $N$ distinct roots if and only if its discriminant $\mathrm{Disc}(P)\neq 0$.
We first present the following lemma.

\begin{lemma}
\label{le-distinct}
Define a Zariski open subset of $\ker(\boldsymbol{U})$ as $\mathcal{C}_{D}\triangleq\{\boldsymbol{c}\in\ker(\boldsymbol{U}) :
\mathrm{Disc}(P)\neq 0\}$.
For generic $\{\alpha_k:k\in[K]\}$, the set $\mathcal{C}_{D}$ is nonempty.
\end{lemma}

\begin{IEEEproof}
The proof is sketched as follows.
The set of vectors $\boldsymbol{\alpha}$ $=$ $(\alpha_0$, $\ldots$, $\alpha_{K-1})$ $\in$ $\mathbb{C}^{K}$
for which $\mathcal{C}_{D}$ $=$ $\emptyset$ defines a proper affine variety in $\mathbb{C}^{K}$,
and therefore has Lebesgue measure zero. Hence, for generic $\{\alpha_{k}:k\in[K]\}$, we have $\mathcal{C}_{D}\neq\emptyset$.
See Appendix C in arXiv \cite{XJM2026CPA} for details.
\end{IEEEproof}

Combining Lemma~\ref{le-C_N}, Lemma~\ref{le-A_K}, and Lemma~\ref{le-distinct},
we consider the intersection
$\mathcal{C}_{N}\cap\mathcal{C}_{A}\cap\mathcal{C}_{D}$.

\begin{lemma}
\label{le-generic-intersection}
For fixed $\{\alpha_k:k\in[K]\}$,
if $\mathcal{C}_{N}$, $\mathcal{C}_{A}$, and $\mathcal{C}_{D}$ are nonempty,
then $\mathcal{C}_{N}\cap\mathcal{C}_{A}\cap\mathcal{C}_{D}$
is a nonempty Zariski open and dense subset of $\ker(\boldsymbol{U})$.
\end{lemma}

\begin{IEEEproof}
Since $\ker(\boldsymbol{U})$ is an irreducible affine variety, and the Zariski open subsets $\mathcal{C}_{N}$, $\mathcal{C}_{A}$, and $\mathcal{C}_{D}$ are non-empty, the intersection $\mathcal{C}_{N}\cap\mathcal{C}_{A}\cap\mathcal{C}_{D}$ is nonempty and Zariski open in $\ker(\boldsymbol{U})$.
Moreover, any nonempty Zariski open subset of an irreducible affine variety is dense.
Hence, $\mathcal{C}_{N}\cap\mathcal{C}_{A}\cap\mathcal{C}_{D}$ is dense in $\ker(\boldsymbol{U})$.
This completes the proof of Lemma~\ref{le-generic-intersection}.
\end{IEEEproof}

From Lemma~\ref{le-generic-intersection}, for generic $\{\alpha_{k}:k\in[K]\}$, all but a Lebesgue measure-zero subset of vectors
$\boldsymbol{c}\in\ker(\boldsymbol{U})$ satisfy the following properties:
1) $c_{N}\neq 0$, so that the normalized polynomial
$P(z)=  \frac{c_{0}}{c_{N}}+\frac{c_{1}}{c_{N}}z+\ldots+z^{N}$ is obtained and satisfies orthogonality conditions;
2) $[\boldsymbol{A}\boldsymbol{c}]_{k}\neq 0$ for all $k\in[K]$, ensuring that $\{\alpha_{k}:k\in[K]\}$ does not contain any root of $P(z)$;
and 3) that the $N$ roots of $P(z)$ are distinct.

This completes the proof of the sufficiency of Theorem~\ref{th-bound-case1}, and hence completes the proof of Theorem~\ref{th-bound-case1}.

\section{Conclusion}
This paper established the fundamental limits of CPA in distributed computing systems.
Unlike existing polynomial coded computing schemes based on individual decoding, we showed that exact recovery of a weighted aggregation of polynomial evaluations can be achieved without reconstructing all individual polynomial evaluations.
We derived a necessary and sufficient condition for the feasibility of CPA in the regime where individual decoding is infeasible.
Based on this characterization, we determined the minimum number of responses required for feasible CPA for generic data points.
Specifically, the minimum number of responses equals $\lfloor (K-1)/2 \rfloor + 1$ for linear computations and $(d-1)(K-1)+1$ for polynomial degree $d \ge 2$, both of which are strictly smaller than the $d(K-1)+1$ responses required by individual decoding.
In addition to the converse, we provided explicit constructions that achieve these limits.

\appendices

\section{Proof of Lemma \ref{le-C_N}}
\label{appendix_le-CN}
We first show that $\mathcal{C}_{N}\neq\emptyset$ if and only if $\boldsymbol{u}_{N} \in \operatorname{span}\{ \boldsymbol{u}_{0}, \boldsymbol{u}_{1},\ldots, \boldsymbol{u}_{N-1}\}$, where $\boldsymbol u_n$ denotes the $n$-th column of $\boldsymbol U$.
If $\mathcal{C}_{N}\neq\emptyset$, then there exists $\boldsymbol c=(c_0,\dots,c_N)^{\mathsf T}\in\ker(\boldsymbol U)$ with $c_N\neq 0$.
From $\boldsymbol U\boldsymbol c=\boldsymbol 0$, we have $\sum_{n=0}^{N} c_n \boldsymbol u_n=\boldsymbol 0$,
which implies $c_N \boldsymbol u_N=-\sum_{n=0}^{N-1} c_n \boldsymbol u_n$.
Since $c_N\neq 0$, dividing both sides by $c_N$ yields $\boldsymbol u_N=\sum_{n=0}^{N-1}\Big(-\frac{c_n}{c_N}\Big)\boldsymbol u_n$,
and hence $\boldsymbol{u}_{N}\in \operatorname{span}\{\boldsymbol u_0,\dots,\boldsymbol u_{N-1}\}$.
Conversely, if $\boldsymbol u_N=\sum_{n=0}^{N-1} e_n \boldsymbol u_n$ for some $(e_0,\dots,e_{N-1})\in\mathbb C^N$, we define
$\boldsymbol c=(e_0,\dots,e_{N-1},-1)^{\mathsf T}$.
Then $\boldsymbol U\boldsymbol c=\sum_{n=0}^{N-1} e_n \boldsymbol u_n-\boldsymbol u_N=\boldsymbol 0$, which implies $\boldsymbol c\in\ker(\boldsymbol U)$ and $c_N=-1\neq 0$. 

Hence, it suffices to show that $\boldsymbol{u}_{N} \in$ $\operatorname{span}\{ \boldsymbol{u}_{0}$, $\boldsymbol{u}_{1}$, $\ldots$, $\boldsymbol{u}_{N-1}\}$ holds for generic $\boldsymbol{\alpha}\in\mathbb C^{K}$.
From  $N\ge C$, we consider the first $C$ columns of $\boldsymbol U$, denoted by $\boldsymbol U' \triangleq$ $[\boldsymbol u_0\ \boldsymbol u_1\ \ldots\ \boldsymbol u_{C-1}] \in\mathbb C^{C\times C}$.
It suffices to show that $\operatorname{rank}(\boldsymbol U')=C$, or equivalently
$\det(\boldsymbol U')\neq 0$, since this implies $\operatorname{span}\{\boldsymbol{u}_0$, $\dots$, $\boldsymbol{u}_{C-1}\}$ $=$ $\mathbb C^{C}$,
and hence $\boldsymbol u_N$ $\in$ $\operatorname{span}\{\boldsymbol u_0$, $\ldots$, $\boldsymbol u_{N-1}\}$. In the following, we will show that the collection of $\boldsymbol{\alpha}$ that leads to $\det(\boldsymbol U') = 0$ is  a strict affine variety in $\mathbb{C}^{K}$, and hence has Lebesgue measure zero.
It follows that for generic $\{\alpha_{k}: k\in [K]\}$,
$\det(\boldsymbol U')\neq 0$ and $\operatorname{rank}(\boldsymbol U')=C$.
Thus, $\boldsymbol u_N$ $\in$ $\operatorname{span}\{\boldsymbol u_0$, $\ldots$, $\boldsymbol{u}_{C-1}\}$ $\subseteq$ $\operatorname{span}\{\boldsymbol u_0$, $\ldots$, $\boldsymbol u_{N-1}\}$, and $\mathcal C_N\neq\emptyset$ for generic $\{\alpha_{k}: k\in [K]\}$.

In detail, we let  $\boldsymbol{A}_{[:,\mathcal{S}]}$ denotes the sub-matrix of matrix $\boldsymbol{A}$ consisting of columns indexed by $\mathcal{S}$, and $\boldsymbol{A}_{[\mathcal{S},:]}$ denotes the sub-matrix consisting of rows indexed by $\mathcal{S}$.   From $\boldsymbol U=\boldsymbol V\operatorname{diag}(\boldsymbol w)\boldsymbol A$, we have $\boldsymbol U'=\boldsymbol V\operatorname{diag}(\boldsymbol w)\boldsymbol A_{[:,[C]]}$.
From the definitions of $\boldsymbol{V}$ and $\boldsymbol{A}$, we have
$\boldsymbol V=\boldsymbol A_{[:,[C]]}^{\mathsf T}$, and thus $\boldsymbol U' = \boldsymbol A_{[:,[C]]}^{\mathsf T}\operatorname{diag}(\boldsymbol w)\boldsymbol A_{[:,[C]]}.$
Applying the Cauchy--Binet formula to
\begin{equation*}
\boldsymbol X\triangleq\boldsymbol A_{[:,[C]]}^{\mathsf T}\operatorname{diag}(\boldsymbol w)\in\mathbb C^{C\times K}, \ 
\boldsymbol Y\triangleq\boldsymbol A_{[:,[C]]}\in\mathbb C^{K\times C},
\end{equation*}
we obtain
\begin{equation}
\label{eq:CB_detUprime}
\det(\boldsymbol U')
=
\sum_{\mathcal S\subseteq[K],\,|\mathcal S|=C}
\det\!\big(\boldsymbol{X}_{[:,\mathcal S]}\big)\,
\det\!\big(\boldsymbol{Y}_{[\mathcal S,:]}\big).
\end{equation}

For each $\mathcal S\subseteq[K]$ with $|\mathcal S|=C$, we have $\boldsymbol{X}_{[:,\mathcal S]}$ $=$ $\boldsymbol{A}_{[\mathcal S,[C]]}^{\mathsf T} $ $ \operatorname{diag}(\boldsymbol {w}_{\mathcal S})$ and $ \boldsymbol{Y}_{[\mathcal S,:]}$ $=$ $\boldsymbol A_{[\mathcal S,[C]]}$,
where $\boldsymbol w_{\mathcal S}\triangleq(w_{k})_{k\in\mathcal S}$.
Therefore, $\det\!\big(\boldsymbol{X}_{[:,\mathcal S]}\big) = \det(\boldsymbol A_{[\mathcal S,[C]]}^{\mathsf T})
\det(\operatorname{diag}(\boldsymbol w_{\mathcal S})) = \Big(\prod_{k\in\mathcal S} w_{k}\Big) \det(\boldsymbol A_{[\mathcal S,[C]]})$, 
and \eqref{eq:CB_detUprime} becomes 
$$\det(\boldsymbol U') = \sum_{\mathcal S\subseteq[K],\,|\mathcal S|=C}
\Big(\prod_{k\in\mathcal S} w_{k}\Big) \det\!\big(\boldsymbol A_{[\mathcal S,[C]]}\big)^2.$$
Each $\boldsymbol A_{[\mathcal S,[C]]}$ is a $C\times C$ Vandermonde matrix built from
$\{\alpha_k:k\in\mathcal S\}$, and hence its determinant is a nonzero polynomial
whenever the $\alpha_k$'s are distinct.
Consequently, $\det(\boldsymbol U')$ is a nonzero polynomial in $\{\alpha_{k}: k\in [K]\}$.
Therefore, the set $ \{ \boldsymbol{\alpha} \in\mathbb C^K:$ $\det(\boldsymbol U')=0\}$
is a strict affine variety in $\mathbb C^K$.  From the discussion above, we conclude that  $\mathcal C_N\neq\emptyset$ for generic $\{\alpha_{k}: k\in [K]\}$.

\section{Proof of Lemma \ref{le-A_K}}
\label{appendix-le-Ak}
\subsection{$\mathcal A_k \neq \emptyset$ for Generic $\{\alpha_k:k\in[K]\}$}
Assume that, for contradiction,
\begin{equation}
    \label{eq-assume-ak}
    [\boldsymbol A \boldsymbol c]_k = 0
\quad \text{for all } \boldsymbol c \in \ker(\boldsymbol U).
\end{equation}
Then the $k$-th row of $\boldsymbol A$, denoted by $\boldsymbol a_k$,
lies in the row space of $\boldsymbol U$, i.e., $\boldsymbol a_k \in  \operatorname{span}\{\boldsymbol u^0,\boldsymbol u^1,\ldots,\boldsymbol u^{C-1}\}$,
where $\boldsymbol u^i$ denotes the $i$-th row of $\boldsymbol U$. 
We consider two cases depending on $d$.

\subsubsection{When $d\ge 2$}
From assumption, there exist coefficients $\lambda_0,\ldots,\lambda_{C-1}$ such that $\boldsymbol a_k $ $=\sum_{i=0}^{C-1} \lambda_i\boldsymbol u^i$. 
Using the factorization $\boldsymbol U$ $=$ $\boldsymbol V$ $\operatorname{diag}(\boldsymbol w)$ $\boldsymbol A$,
let $\boldsymbol v^i$ denote the $i$-th row of $\boldsymbol V$, whose $l$-th entry is
$v^i[l]=\alpha_l^{\,i}$.
Then $\boldsymbol u^i$ $=$ $\boldsymbol v^i$ $ \operatorname{diag}(\boldsymbol w)$ $\boldsymbol A$ $=$ $\sum_{l=0}^{K-1} \alpha_l^{\,i} w_l \boldsymbol a_l$.
Substituting this into the expression for $\boldsymbol a_k$ and exchanging the order of summation yields $\boldsymbol a_k =\sum_{l=0}^{K-1} w_l
\Big(\sum_{i=0}^{C-1}\lambda_i \alpha_l^{\,i}\Big)\boldsymbol a_l$.
Define the degree-$(C-1)$ polynomial $Q(z)\triangleq\sum_{i=0}^{C-1}\lambda_i z^i$.
Then $\boldsymbol a_k=\sum_{l=0}^{K-1} w_l Q(\alpha_l)\boldsymbol a_l$.
Let $x_l\triangleq w_l Q(\alpha_l)$ for $l\in[K]$.
Comparing the $n$-th coordinate of $\boldsymbol a_k$ gives
\begin{equation}
\label{eq-moment-x}
\sum_{l=0}^{K-1} x_l \alpha_l^n = \alpha_k^n,\qquad n \in [N+1] .
\end{equation}
Under the sufficiency condition of Theorem~\ref{th-bound-case1} for $d \geq 2$, i.e., $N\ge K$, we restricting \eqref{eq-moment-x} to $n=0,1,\ldots,K-1$ yields the Vandermonde system $\boldsymbol V_K \boldsymbol x = \boldsymbol b_k$,
where $(\boldsymbol V_K)_{n,l}=\alpha_l^n$ and $\boldsymbol b_k=(\alpha_k^0,\alpha_k^1,\ldots,\alpha_k^{K-1})^{\mathsf T}$ is the $k$-th column of $\boldsymbol V_K$.
Since the $\{\alpha_l\}$ are distinct, the Vandermonde matrix $\boldsymbol V_K$ is invertible.
Moreover, $\boldsymbol b_k$ is exactly the $k$-th column of $\boldsymbol V_K$. Hence the unique solution to $\boldsymbol V_K \boldsymbol x=\boldsymbol b_k$ is $\boldsymbol x=\boldsymbol e_k$.
Equivalently, $w_l Q(\alpha_l)=0 $ for all $l\neq k$, and  $w_k Q(\alpha_k)=1$.
Thus $Q(\alpha_l)=0$ for all $l\neq k$. However, $\deg(Q)\le C-1$ and $C<K$, so $Q$ cannot have $K-1$ distinct zeros for generic
distinct $\{\alpha_l\}$. This contradiction shows that $\boldsymbol a_k\notin \operatorname{span}\{\boldsymbol u^0,\boldsymbol u^1,\ldots,\boldsymbol u^{C-1}\}$ for generic $\{\alpha_k:k\in[K]\}$.

\subsubsection{When $d=1$}
Condition \eqref{eq-assume-ak} is equivalent to $\boldsymbol a_k \boldsymbol c = 0$ for all $\boldsymbol c \in \ker(\boldsymbol U)$.
This implies 
\begin{equation}
\boldsymbol a_k \in (\ker(\boldsymbol U))^\perp = \operatorname{col}(\boldsymbol U^{\mathsf T}).
\label{eq:ak_in_colUT}
\end{equation}

Next, we show that \eqref{eq:ak_in_colUT} cannot hold for generic
$\{\alpha_k:k\in[K]\}$.
Since $\boldsymbol U \in \mathbb C^{C\times (N+1)}$, we have $\dim \operatorname{col}(\boldsymbol U^{\mathsf T})$ $=$ $\operatorname{rank}(\boldsymbol U)$ $\le$ $C < N+1$.
On the other hand, $\boldsymbol a_k=$ $(1$, $\alpha_k$, $\alpha_k^2$, $\ldots$, $\alpha_k^N)$
depends on $\alpha_k$. The condition $\boldsymbol a_k \in \operatorname{col}(\boldsymbol U^{\mathsf T})$
therefore imposes a nontrivial algebraic constraint on
$\{\alpha_k:k\in[K]\}$, and hence holds only on a strict affine variety
of $\mathbb C^K$.
Consequently, for generic $\{\alpha_k:k\in[K]\}$,
\eqref{eq:ak_in_colUT} does not hold, which contradicts
\eqref{eq-assume-ak}.

Combining the two cases, we conclude that for generic $\{\alpha_k:k\in[K]\}$,
there exists $\boldsymbol c\in\ker(\boldsymbol U)$ such that
$[\boldsymbol A\boldsymbol c]_k\neq 0$.
Hence $\mathcal A_k\neq\emptyset$.

\subsection{$ C_A \neq \emptyset $ for Generic $\{\alpha_k:k\in[K]\}$ }
Since $\ker(\boldsymbol{U})$ is a linear subspace and hence an irreducible affine variety, any finite intersection of nonempty Zariski open subsets of $\ker(\boldsymbol{U})$ is nonempty.
Hence, $\mathcal{C}_{A}\neq\emptyset$.

\section{Proof of Lemma~\ref{le-distinct}}
\label{appendix-le-Cd}
Since $P(z)=P_{\mathrm{form}}(z)/c_N$ with $c_N\neq 0$, the discriminants
$\mathrm{Disc}(P)$ and $\mathrm{Disc}(P_{\mathrm{form}})$ differ only by a nonzero multiplicative constant. Hence,
$\mathrm{Disc}(P)\neq 0$ if and only if $\mathrm{Disc}(P_{\mathrm{form}})\neq 0$.
Recall that $P_{\mathrm{form}}(z)=\sum_{n=0}^{N} c_n z^n$ and that $\mathrm{Disc}(P_{\mathrm{form}})$
is a polynomial in the coefficients $\boldsymbol{c}$.
Moreover, $\mathrm{Disc}(P_{\mathrm{form}})$ is not identically zero, since there exist polynomials with distinct roots.
Hence,
$\mathcal{O}\triangleq\{\boldsymbol{c}\in\mathbb{C}^{N+1}:\mathrm{Disc}(P_{\mathrm{form}})\neq 0\}$
is a nonempty Zariski open subset of $\mathbb{C}^{N+1}$.

Fix $\{\alpha_k:k\in[K]\}$ and the corresponding matrix $\boldsymbol{U}$.
The set
$\mathcal{C}_D=\{\boldsymbol{c}\in\ker(\boldsymbol{U}):\mathrm{Disc}(P_{\mathrm{form}})\neq 0\}$
is the intersection of $\ker(\boldsymbol{U})$ with the Zariski open set $\mathcal{O}$,
and is therefore Zariski open in $\ker(\boldsymbol{U})$.
Moreover, $\mathcal{C}_D$ is empty if and only if
$\mathrm{Disc}(P_{\mathrm{form}})$ vanishes identically on $\ker(\boldsymbol{U})$.

Since the entries of $\boldsymbol{U}$ depend polynomially on $\boldsymbol{\alpha}$,
the set
\begin{align*}
    \mathcal{B} & \triangleq\{\boldsymbol{\alpha}\in\mathbb{C}^{K}:\mathcal{C}_D=\emptyset\} \\
& =\{\boldsymbol{\alpha}:\mathrm{Disc}(P_{\mathrm{form}})\text{ vanishes identically on }\ker(\boldsymbol{U}(\boldsymbol{\alpha}))\}
\end{align*}
is an affine variety of $\mathbb{C}^{K}$. Moreover, $\mathcal{B}$ is proper because there exists at least one choice of
$\boldsymbol{\alpha}$ for which $\mathcal{C}_D \neq \emptyset$.
Consequently, for generic $\{\alpha_k:k\in[K]\}$, the set $\mathcal{C}_D$ is nonempty.

\bibliographystyle{IEEEtran}
\bibliography{reference}

\end{document}